\documentstyle[aps,pre,epsfig]{revtex}
 \newcommand {\bi} {\bibitem}
 \newcommand {\be} {\begin{equation}}
\newcommand {\bea} {\begin{eqnarray} \nonumber }
\newcommand {\ee} {\end{equation}}
\newcommand {\eea} {\end{eqnarray}}
 \newcommand {\eps} {\epsilon}

\newcommand {\la} {\langle}
\newcommand {\ra} {\rangle}

\newcommand {\lam} {\lambda}
\newcommand {\psih} {{\hat \psi}}
\newcommand {\chih} {{\hat \chi}}

\newcommand {\ch} {{\hat c}}
\newcommand {\zh} {{\hat z}}
\newcommand {\cZ} {{\cal Z}}
\newcommand {\tf} {{\tilde f}}

\def\(({\left(}
\def\)){\right)}
\def\[[{\left[}
\def\]]{\right]}

\begin{document}
\draft


\title{Spectra of Euclidean Random Matrices}
\author{M. M\'ezard$^{(1,2)}$, G. Parisi$^{(3)}$, A. Zee$^{(1)}$}
\address{
$^{(1)}$ Institute for Theoretical Physics,\\
 University of California Santa Barbara, CA 93106-4030 (USA)\\
$^{(2)}$ CNRS, Laboratoire de Physique Th\'eorique de l'ENS,\\
 24 rue Lhomond, 75231 Paris
  (France)\\
$^{(3)}$ Dipartimento di Fisica and Sezione INFN,\\
Universit\`a di Roma ``La Sapienza'',
Piazzale Aldo Moro 2,
I-00185 Rome (Italy)
}
\date{\today}
\maketitle
\abstract{
We study the spectrum of a random matrix, whose elements depend on 
the Euclidean distance between points randomly distributed in  space. 
This problem is widely studied in the context of the  Instantaneous Normal 
Modes of fluids and  is particularly relevant at the glass transition. 
We introduce a systematic study of this problem through its representation 
by a field theory. In this way we can easily construct a high density 
expansion, which can be resummed producing an approximation to the spectrum
similar to the Coherent Potential Approximation for  disordered systems.
}

\section{introduction}
The theory of random matrices has found applications in many branches of physics
\cite{RMT}.
The most developed theory concerns matrices where the matrix elements are either
independent random variables, as in Gaussian ensembles, or are taken with a
statistical distribution which is invariant under some symmetry group. However
in many physical applications, from vibration spectra of glasses
\cite{glasses,MPg} to instantaneous normal modes in liquids \cite{liq},
electron hopping in amorphous semiconductors \cite{glasses}
  or combinatorial optimization problems
\cite{MP},
one needs to compute the spectrum and the eigenstate properties
 of some random matrices which are of a different type: The disorder
is due to the random positions of $N$ points, and the matrix elements
are given by a deterministic function of the distances between the points.
We shall call these matrices Euclidean Random Matrices(ERM).

Specifically
we want to study the following mathematical problems:
Consider $N$ points in a volume $V$ of a $d$-dimensional Euclidean space.
For a given sample, characterized by the positions $x_i$ of the
$N$ points ($ i \in \{ 1...N\} $), we
want to study the properties of the $N \times N$ random matrices $M$
defined as
\be
M_{ij}= f(x_i-x_j) -u \delta_{ij} \sum_{k } f(x_i-x_k) \ ,
\label{pb}
\ee
where $u$ is a real parameter which enables us to interpolate between the
two most interesting cases $u=0,1$.
The case where $u=0$ is the simplest mathematical problem with Euclidean-correlated
matrix elements, the case where $u=1$ is the natural problem which appears when
studying for instance vibration modes of an amorphous solid,
instantaneous normal modes of a liquid,
or random master equations. The main difference is that when $u=1$ the matrix
$M$ has fluctuating diagonal terms, tailored in such a way that
$\sum_j M_{ij}=0$: the vector
with all components equal to one is an eigenvector with zero
eigenvalue, which expresses global
translation  invariance.

To fully specify the problem one needs to characterize the probability
distribution of the random points $x_i$, as well as the function $f(x)$.
In this paper we shall concentrate on the case where the points are uniformly
distributed in a cubic box of size $L=V^{1/d}$, without correlations. In many
applications one will have to generalize the problem to the case where the
$x_i$'s are correlated, including a short distance repulsion. This is a much
more complicated problem which we shall not address here. As for the function
$f(x)$, we shall assume that it depends only on the distance $|x|$ and that
it decays fast enough at large argument. In particular we shall
assume that the Fourier transform $\tilde f(k) \equiv \int dx e^{ik.x} f(x)$
is a well defined function at all $k$.

The general problem of ERM theory is to understand the statistical
properties of the eigenvalues  and the corresponding eigenvectors
of $M$ in the large $N$ limit (taken at fixed
density $\rho=N/V$). This is certainly a very rich problem. In particular
one can expect that the eigenvectors  will exhibit in dimension $d \ge 3$
some localized and delocalized regimes separated by a mobility edge
\cite{loc_rev}. Here
we wish to keep to the much simpler question of the computation of the spectrum
of $M$.
We shall  develop  a field theory for this problem,
check it at high and low densities, and use
a Hartree type method to approximate the spectrum in the $u=0$ case.

\section{Field theory}
The spectrum can be computed from the resolvent:
\be
R(z)={1 \over N}\overline{ Tr {1 \over z-M}} \ ,
\label{res_def}
\ee
where the overline denotes the average over the positions $x_i$.
It is possible to write down a field theory using a replica
approach. We shall compute $\Xi_N \equiv \overline{ \det (z-M) ^{-n/2}} $,
and deduce from it the resolvent by
using the replica limit $n \to 0$.
It is easy to show that
one can write $\Xi_N$ as a partition function
over replicated fields $\phi_i^a,$ where $i\in\{ 1...N\} , \ a\in \{1...n\} $:
\bea
\Xi_N=&&
\int \prod_{i=1}^N {dx_i \over V} \int \prod_{i=1}^N \prod_{a=1}^n
d \phi_i^a \\
&&\exp\((-{z \over 2} \sum_{i,a} (\phi_i^a)^2+ {1 \over 2} \sum_{i,j,a}
f(x_i-x_j) \phi_i^a \phi_j^a
- {u \over 2} \sum_{i,a}
 (\phi_i^a)^2 \sum_j f(x_i-x_j) \label{PAFUNC}
\)) \ .
\eea
Let us introduce the bosonic fields $\psi_a(x)=\sum_{i=1}^N \phi_i^a \delta(x-x_i)$
and $\chi(x)=  \sum_{i,a} (\phi_i^a)^2 \delta(x-x_i)$,
together with their respective  Lagrange multiplier fields $\hat \psi_a(x)$ and
$\hat \chi (x)$. One can integrate out
the  $\phi$  variables, leading to the following field theory for $\Xi_N$:
\be
\Xi_N= \int D[\psi_a,\hat \psi_a,\chi,\chih] A^N
\exp\((S_0\))
\ee
where
\bea
S_0&=&i \sum_a \int dx\  \psih_a(x) \psi_a(x) +{1 \over 2} \int dx
\ \chih(x) \chi(x)
+ {1 \over 2} \sum_a
\int dx dy \ \psi^a(x) f(x-y) \psi^a(y)  \ ,\\
A&=& \int {dx \over V} \(({2\pi \over z+\chih(x)}\))^{n/2}
\exp\[[-{1 \over 2} u \int dy f(x-y) \chi(y)-{1 \over 2(z+\chih(x))}
 \sum_{a} \psih^a(x)^2 \]] \ .
\eea
It is convenient to go to a grand canonical formulation of the disorder: we
consider an ensemble of samples with varying number of points, and compute the grand canonical
partition function
${\cZ}(\alpha) \equiv \sum_{N=0}^\infty \Xi_N \alpha^N /N!$
which is equal to:
\be
\cZ= \int D[\psi_a,\hat \psi_a,\chi,\chih]  \exp\((S_0 + S_1\)) \ \  ; \  \
 S_1 \equiv \alpha A \ .
\label{cZ1}
\ee
Notice that we can also integrate out the $\psi$ field thus replacing
$S_0$ by $S_0'$, where
\be
S_0'=-{n \over 2} tr \log f+{1 \over 2} \int dx \ \chih(x) \chi(x)
+ {1 \over 2} \sum_a
\int dx dy \ \psih^a(x) f^{-1}(x-y) \psih^a(y),
\label{cZ2}
\ee
where $\log f$ is the logarithm of $f$ considered as an integral operator 
and $f^{-1}$ is the integral operator which is the inverse of $f$.

The  expression (\ref{cZ1}) is our basic field theory representation.
 We shall denote by brackets
the expectation value of any observable  with the
action $S_0+S_1$.
As usual with the replica method we have traded the disorder for an interacting replicated
system. The basic properties of the field theory are related
to the properties of the original problem in a straightforward way.
The average number of particles is related
to $\alpha$ through $N=\alpha \la A \ra$,
so that one gets $\alpha=N$ in the $n \to 0$ limit.
 From the generalized partition function $\cZ$, one can get the resolvent $R(z)$
through:
\be
R(z)=- \lim_{n \to 0} {2 \over n N} {\partial \log \cZ \over \partial z}
\label{RfromZ}
\ee

\section{High density expansion}
Let us first show how this field theory can be used to derive a high density
expansion. We shall rescale $z$ as $z=\rho \hat z$, and rescale the $\chi$ fields
as: $\chih(x)= \rho \hat c (x)$, $\chi(x)=c(x)/\rho$. The interacting part
of the action, $S_1$, can then be expanded as:
\be
S_1 \simeq \rho \int dx \mu(x)-{u \over 2} \int dx dy \mu(x) f(x,y) c(y)
+ {u^2 \over 8 \rho} \int dx dy \ c(x) h(x,y) c(y)
-{1 \over 2} \int dx {\sum_a \psih_a(x)^2 \over \zh +\ch(x)}  \ ,
\ee
where we have introduced:
\be
\mu(x) \equiv \(({2 \pi \over \rho(\zh +\ch(x))}\))^{-n/2}   \ \ ; \ \
h(x,y) \equiv \int dr f(x,r) \mu(r) f(r,y) \ .
\ee
Performing the quadratic  $c$ functional integral, one finds (here
and in the following we drop all the irrelevant -$\hat z$ independent-
 constants):
\bea
\cZ&\propto&
\int D[\psih_a,\ch]
\\&&\exp\[[ \rho \int dx \mu(x)
- {1 \over 2} \int dx dy \ \eps(x) h^{-1}(x,y) \eps(y)
- {1 \over 2} \sum_a
\int dx dy \ \psih^a(x) G^{-1}(x,y) \psih^a(y) \]] \ ,
\eea
where $G^{-1}(x,y)=\delta(x-y) /(\zh+\ch(x))-f^{-1}(x,y)$ is the
propagator of the $\psih$ field,
and the field $\eps(x)$ is defined as:
\be
\eps(x)= {\sqrt{\rho} \over u}\((\ch(x)-u \int dy f(x-y) \mu(y)\))
= {\sqrt{\rho} \over u}
\(( \ch(x)-u \tf(0)- {u n \over 2} \int dy f(x-y) \log(\zh+\ch(y)) \)) \ .
\ee
We then change variables from $\ch$ to $\eps$, which involves a Jacobian which
is easily computed to order $n$.
The $\psih$ integral is  Gaussian, and to leading order
in $\rho$, the $\eps$ field can be
neglected in $G$, as well as in $h$.
The only term in which $\eps$ is relevant turns out to be $\int dx \mu(x)$, which
is simplified by an expansion to order $\eps^2$. Gathering the various
contributions one gets
\bea
\log\cZ \simeq &-&{nN \over 2}
\((
\log(\hat z + u \tf(0))
-{u^2 \over 2 \rho} {\int (dk) \tf(k)^2 \over (\zh+u\tf(0))^2}
+ {1 \over \rho}
{f(0) \over \zh + u \tf(0)}
\right.
\\
&-&
\left.
{1 \over \rho} \int (dk) \log\[[ {\zh+u \tf(0)-\tf(k) \over \zh+u \tf(0)} \]]
+O\(( {1 \over \rho^2} \))
\)) \ ,
\eea
where the first two terms come from the expansion of $\int dx \mu(x)$,
the third one is the Jacobian, and the fourth one is the contribution
of the determinant from the $\psih$ fluctuations.

This gives for the resolvent:
\bea
R(z)={1 \over \rho} {1 \over \hat z +u \tf(0)}&+&
{1 \over \rho^2} \(( \int (dk)
{\tf(k) \over ( \hat z +u \tf(0)-\tf(k))(\hat z+ u\tf(0))}\right.
\\
&-&
\left.
{uf(0) \over
(\zh+u\tf(0))^2} + { u^2 \int dr f(r)^2 \over (\zh+u\tf(0))^3} \))
+O\(({1 \over \rho^3}\)) \ .
\label{Rleading}
\eea
This result is valid in the large $\rho$ limit, whenever the resolvent parameter
$z$ is scaled as $z=\rho \zh$.
It can be checked directly. We consider the
original form (\ref{res_def}) of the resolvent and expand it in a $1/z$ series.
At order $1/z^{p+1}$, the first two leading terms in $\rho$ are:
\bea
\rho^p \((-u \tf(0)\))^p+p (1-u) f(0) \rho^{p-1} \((-u \tf(0)\))^{p-1}
&+&\rho^{p-1}  \sum_{p'=2}^p C_p^{p'}
\int dk [-u \tf(0)]^{p-p'} [\tf(k)]^{p'} \\
&+& \rho^{p-1} u^2 C_p^2 \((-u\tf(0)\))^{p-2} \int dr f(r)^2 \ ,
\eea
where $C_p^{p'}=\Gamma(p+1)/(\Gamma(p'+1)\Gamma(p-p'+1))$.
One can see  that in the large $\rho$ limit the series simplifies if we first
scale $z$ proportionally to $\rho$. Writing $z=\rho \hat z$, we can
resum the leading terms in the series, which gives back the same
resolvent as (\ref{Rleading}).

One can study with this method the eigenvalue
density for eigenvalues $|\lam| \sim O(\rho)$, by taking $z=\lam + i \eta$
and computing the imaginary part of the resolvent in the small $\eta$
limit. For $\rho \to \infty$ one would get the trivial result
for the eigenvalue density$ P(\lam)=\delta(\lam+\rho u\tf(0))$. Including
the leading large $\rho$ correction which we have just computed,
we find that $P(\lam)$ develops,
away from the peak at $\lam \sim- \rho  u\tf(0)$, a component of the form:
\be
P(\lam) \sim {1 \over \rho} \int (dk) \delta ( \lam-\rho(\tf(k)-u\tf(0))) \ .
\label{spec_highrho}
\ee
The result (\ref{spec_highrho}) is the one that one could guess using the following
simple argument: introducing
$\phi_j=\exp(ikx_j)$, one gets:
\be
\sum_j M_{ij} \phi_j = \lam_i \phi_i \ \ , \ \
\lam_i= \sum_j e^{-ik(x_i-x_j)} f(x_i-x_j)-u \sum_m f(x_i-x_m) \ .
\label{eveq}
\ee
In the large density regime it is reasonable to use the central limit theorem
to approximate the above sum, which would show that $\phi$ is near to an eigenvector,
with eigenvalue $\rho (\tf(k)-u \tf(0))$. Notice that for this argument
to hold, the discrete sum giving $\lam_i$ in (\ref{eveq}) must sample correctly
the continuous integral. This will be the case only when
the density $\rho$ is large enough that the phase $-ik(x_i-x_j)$ doesn't
oscillate too much from one point $x_j$ to a neighbouring one.
This imposes that the spatial frequency $|k|$ be small enough: $|k| \ll \rho^{1/d}$.
This same condition is present in the field theory derivation. We assume that $\tf(k)$
decreases at large $k$, and we call $k_M$ the typical range of $k$ below which $\tf(k)$
can be neglected.
Let us consider the corrections
of order  $\rho^{-2}$ in eq. (\ref{Rleading}). It is clear that,
provided $\hat z$ is away from $-u \tf(0)$, the ratio of the correction
term to the leading one is of order
$k_{M}^{d}\rho^{-1}$, and the condition that the correction be small
 is just identical to the previous one. The large density corrections near
 to the peak $\lam \sim- \rho  u\tf(0)$ cannot be studied with this method.

\section{Low density expansion}
The low density expansion is also easily performed from the field theoretic
representation. Since the interaction $\alpha A$ is proportional
to $\rho$ in the $n \to 0$ limit, we can expand in powers of $\alpha$:
\bea
\cZ&=& \int D[\hat \psi_a,\chi,\chih] e^{S_0'}
\\
&&\[[1 +
\alpha \int {d x_0 \over V} \(({2 \pi \over z+\chih(x_0)}\))^{n/2}
\exp\((-{u \over 2} \int dy f(x_0-y) \chi(y)-{1 \over 2(z+\chih(x_0))}
\sum_a \psih^a(x_0)^2\))+...\]]
\label{cZlow}
\eea
At first order we can perform the $\chi$ integral, which fixes $\chih(x)=  u f(x-x_0)$
and we get:
\be
\cZ= 1 +\rho \(({2\pi \over z}\))^{n/2} \(({\det K_1}\))^{-n/2}
\ee
where $K_1$ is the operator:
$K_1(x,y)=\delta(x-y)- \delta (x-x_0) \delta(y-x_0) f(0)/(z+u f(0)) $. Using
(\ref{RfromZ}), one finds  $R(z)=1/(z-(1-u)f(0))$, which is obviously
the leading result at very low densities, such that the points are isolated.
At second order, the expansion (\ref{cZlow}) involves a double integral over
points $x_0, x_1$. The $\chi$ integral fixes $\chih(x)=u f(x-x_0)+uf(x-x_1)$.
One needs to study the determinant of the operator $K_2$ defined
as:
\be
K_2=\delta(x-y)- {f(x-y) \over z+\chih(x)}\((\delta(x-x_0)+\delta(x-x_1)\))
\((\delta(y-x_0)+\delta(y-x_1)\)) \ .
\ee
This gives after a simple computation the order $\rho$ correction to the
resolvent:
\be
 {\rho \over 2} \int dr \((
{1 \over z-[(1-u)f(0)+(1-u)f(r)]}+{1 \over z-[(1-u)f(0)-(1+u)f(r)]} -{2 \over
z-(1-u)f(0)} \)) \ .
\label{lowrho}
\ee
This corrects the leading small $\rho$ formula by adding the contribution due
to pairs of points, a distance $r$ apart, which are isolated from all other points.
This expansion can be easily carried out to higher order,
the order $\rho^k$ involving the computation of a $k d$ dimensional integral.

\section{Variational approximation}
In order to try to elaborate a general approximation for the spectrum, which interpolates
between the high and low density limits, we have used a standard Gaussian
variational approximation in the field theory representation, which appears
under various names in the literature, like
the Random Phase Approximation. We have developed it only in the case $u=0$.
The field theory then simplifies since the $\chi$ and $\chih$ fields
can be integrated out. Changing $\psih \to i \psih$, we obtain:
\be
\cZ= \int D[\psih^a] \exp\((S_{u=0}\)) \ ,
\ee
where
\be
S_{u=0} = -{1 \over 2}
\int dx dy
\sum_a \psih^a(x) f^{-1}(x,y) \psih^a(y)
+\rho z^{-n/2} \int dx
\exp \(( {1 \over 2 z} \sum_a \psih^a (x)^2 \))  \ .
\ee
We look for the best
quadratic action
\be
S_v=-(1/2) \sum_{ab} \int dx dy G^{-1}_{ab}(x,y) \psih^a(x) \psih^b(y)
\ee
which approximates the full interacting problem,
using the fact that the variational free
energy $F_v= <S_{u=0}>_v-\log \cZ_v$ should be minimized. (Here $\cZ_v=
\int D[\psih] \exp(S_v)$, and the expectation values $<.>_v$ are meant as
Boltzmann like averages with the measure $\exp(S_v)$). The variational
free energy is easily obtained:
\be
{F_v \over V} = {1 \over 2} Tr f^{-1} G
-\rho z^{-n/2} \int dx_0 \(({det(K) \over det(G)} \))^{1/2}
-{1 \over 2}  Tr \log  G \ ,
\label{Fv}
\ee
where the operator $K$ is defined as:
\be
K^{-1}_{ab}(x,y)=G^{-1}_{ab}(x,y)-{1 \over z} \delta_{ab} \delta(x-x_0)
\delta(y-x_0) \ .
\ee
In (\ref{Fv}), both $G$ and $K$ are considered as operators in $x$ space and
in replica space. Keeping to translation invariant variational actions,
the result is easily expressed in terms of the Fourier transform $G_{ab}(k)$
of the function $G_{ab}(r)$. One finds that:
\be
 {det(K) \over det(G)} = \exp\[[ Tr_n \log \((1-{H \over z}\))  \]] \ ,
\ee
where $H_{ab}=G_{ab}(r=0) = \int (dk) G_{ab}(k)$ is a $n \times n$ matrix, and the
trace $Tr_n$ is a trace in replica space. This gives for the
variational free energy:
\be
{F_v \over V} = {1 \over 2} \int (dk) \  {Tr_n G(k) \over f(k)}
-\mu \exp\((-{1 \over 2} Tr_n \log (1-H/z)\))
-{1 \over 2} \int (dk) Tr_n \log  G(k)
\label{Fv2}
\ee

 We have
found a solution of the stationarity equations,
$\partial F_v/ \partial G_{ab}(x) =0$, of the form $G_{ab}(x)=\delta_{ab}
 G(x)$, where the Fourier transform $\tilde G$ satisfies the self consistency equation:
\be
 \tilde G (k) = { f(k) \over 1-C f(k) } \ \ , \ \
 C={\rho \over z-\int (dk) \tilde G(k)}
\label{Ma}
\ee
In terms of this function, the variational free energy density per replica is:
\be
{F_v \over nV} = {1 \over 2} \int (dk) {\tilde G(k) \over \tilde f(k)}
+{\rho \over 2} \log \(( 1 -{1 \over z} \int (dk) \tilde G(k)\))
-{1 \over 2} \int (dk) \log \tilde G(k)
\label{Mb}
\ee
and the resolvent is obtained from:
\be
R(z)={1 \over z} + {2 \over \rho} {\partial \over \partial z} {F_v \over nV}={C \over \rho}
={1 \over z-\int (dk) \tilde G(k)}
\label{Mc}
\ee

Formulas (\ref{Ma},\ref{Mb},\ref{Mc}) provide a closed set of equations
which allow us to compute the Gaussian variational approximation to the
spectrum for any values of $f$ and the density. In sect. \ref{num}
we shall compare this approximation to some numerical estimates of the spectrum for
various functions $f$ and densities. It is easily checked that the variational
gives the correct spectrum to leading order
in the large $\rho$ expansion (see (\ref{Rleading})) and in the small $\rho$ expansion,
which is not a surprise since we had seen that these leading orders are given by
purely Gaussian theories. Of course it fails to give the exact result beyond leading
order,  but it provides a reasonable interpolation between these extremes.

Another partial resummation of the $\rho$ expansion can also be done in the following
way: if one neglects the triangular-like
correlations between the distances of the points (an approximation which
becomes correct in large dimensions, provided the function $f$
is rescaled properly with dimension), the problem maps
onto that of  a diluted random matrix with independent elements.
This problem can be studied explicitly using the methods of
\cite{Abou,BraRod,FyoMir,BouCiz,BirMon,CavGiaPar}. It
leads to integral equations which can be solved numerically.
The equations one gets correspond to  the first order in the virial
expansion in eq.
(\ref{PAFUNC}), in which one introduces as variational parameter the local
probability distribution
of the field $\phi$. The detailed computations involving this other
approximation scheme are left for future work.

\section{Numerical Simulations}
\label{num}

For a function $f(x)$ which is positive, as well as its Fourier transform
$\tf(k)$ , simple bounds 
on the spectrum can be derived  in the two cases $u=0$ and $u=1$.
In the $u=0$ case, calling $\psi_i$ a normalised eigenvector $\psi_i$ of
the matrix $M$ defined in (\ref{pb}), with eigenvalue $\lambda$, one
has:
\be
\sum_{ij} \psi_i f(x_i-x_j) \psi_j = \lambda \ ,
\ee
and the positivity of the Fourier transform of $f$ implies that $\lambda \ge 0$.
In the $u=1$ case, the eigenvalue equation implies that:
\be
\sum_{j (\ne i)} |M_{ij}| |\psi_j| \ge |\lambda-M_{ii}| |\psi_i| \ .
\ee
Summing over $i$, using the fact that $M_{ij}$  is positive for $i \ne j$,
together with the constraint $\sum_j M_{ij}=0$, one gets:
\be 
\sum_j \(( |M_{jj}| -|\lambda-M_{jj}| \)) | \psi_j| \ge 0
\ee
which implies that there exists at least one index $j$ such
that $|M_{jj}| -|\lambda-M_{jj}|\ge 0$, and therefore  $\lambda \le 0$. 
Furthermore, in this $u=1$ case,
there exists one eigenvector (the uniform one) with zero eigenvalue.

In the $u=0$ case, we have studied numerically the problem in dimension $d=3$
 with the  Gaussian
function $f(x)=(2\pi)^{-3/2}\exp(-x^2/2)$. 
 In this Gaussian case
the high density approximation gives a spectrum 
\be
P(\lambda) \sim {1 \over \rho \pi^2 } {1 \over \lambda}
\(({1 \over 2} \log{\rho \over \lambda}\))^{1/2} \theta(\rho-\lambda)
\label{gauss_highrho}
\ee
Notice that this spectrum is supposed to hold away from the small $\lambda$ peak,
and in fact it is not normalizable at small $\lambda$.

The variational approximation computation described in the previous section
is more involved. From (\ref{Ma},\ref{Mb},\ref{Mc}),
one needs to solve,  given $z=\lambda-i\eps$, the following
equations giving the complex function of $z$, $C(z)$
which we write as $C=a+ib$:
\bea
\lambda&=&\rho {a \over a^2+b^2} + {1 \over 2 \pi^2} \int_0^\infty k^2 dk
{e^{k^2/2}-u \over \((e^{k^2/2}-a\))^2 + b^2} \\
\eps &=& \rho {b \over a^2+b^2} -{b \over 2 \pi^2} \int_0^\infty k^2 dk
{1 \over \((e^{k^2/2}-a\))^2 + b^2} \ .
\label{eqgauss}
\eea
One needs to find a solution in the limit where $\eps \to 0$. We have done this
as follows: For a given value of $a$, we search for the values of $b$ which solve
the second of eqs.(\ref{eqgauss}) with $\eps=0$. Depending on the value of $a$,
there may exist zero, one, or two solutions in $b$. For each such solution
we compute the value of $\lambda$ from the first of qs.(\ref{eqgauss}),
and the corresponding density of states is given by $b/(\rho \pi)$.

\begin{figure}
\centerline{\hbox{
\epsfig{figure=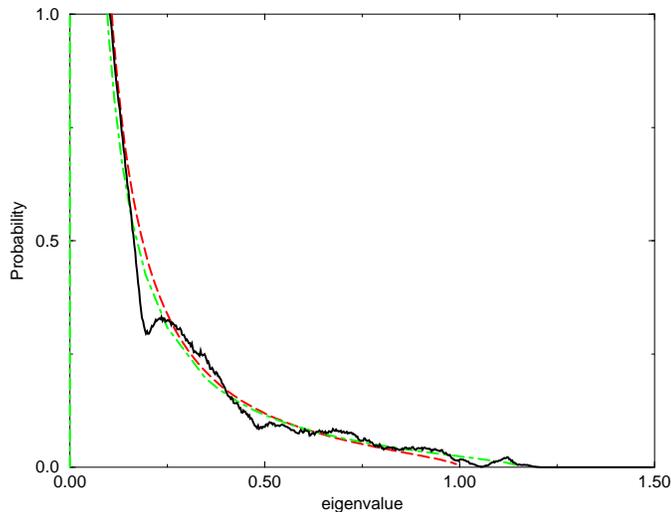,width=8 cm,angle=-90}
}}
\caption{ Density of eigenvalues of a Euclidean Random Matrix in three dimensions,
density $\rho=1$.
The function $f$ is
$f(x)=(2\pi)^{-3/2}\exp(-x^2/2)$, and the matrix is defined from
eq.(\ref{pb}) with $u=0$. The full line is the result of a numerical simulation
with $N=800$ points, averaged over $100$ samples. The dashed (red) line
is the result from the high density expansion. The dash-dotted (green) line is the result 
from the Gaussian variational approximation (RPA) to the field theory.}
\label{fig1}
\end{figure}

\begin{figure}
\centerline{\hbox{
\epsfig{figure=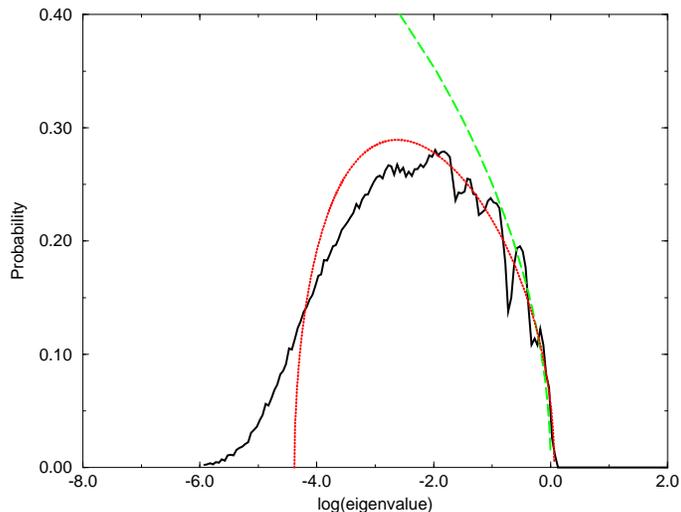,width=8 cm,angle=-90}
}}
\caption{ Density of the logarithm (in base 10) of the
eigenvalues of a Euclidean Random Matrix in three dimensions, density $\rho=1$.
The function $f$ is
$f(x)=(2\pi)^{-3/2}\exp(-x^2/2)$, and the matrix is defined from
eq.(\ref{pb}) with $u=0$. The full line is the result of a numerical simulation
with $N=800$ points, averaged over $100$ samples. The dashed (red) line
is the result from the high density expansion. The dotted (green) line is the result 
from the Gaussian variational approximation (RPA) to the field theory.}
\label{fig2}
\end{figure}

In fig. (\ref{fig1},\ref{fig2}), we plot the obtained spectrum,
averaged over 100 realizations, for $N=800$ points at  density $\rho=1$
(We checked that with $N=600$ points the spectrum is similar). Also
shown are the high density approximation (\ref{gauss_highrho}), and the result from
the variational approximation. We see from fig.(\ref{fig1}) that the 
part of the spectrum $\lambda \in [0.2,1.5]$ 
is rather well reproduced from both approximations, although
the variational method does a better job at matching the upper edge.
On the other hand the probability distribution of the logarithm of the eigenvalues
(fig.\ref{fig2}) makes it clear that the high density approximation is not
valid at small eigenvalues, while the variational approximation gives a
sensible result. One drawback of the variational approximation, though,
is that it always produces sharp bands with a square root singularity,
in contrast to the tails which are seen numerically.

\begin{figure}
\centerline{\hbox{
\epsfig{figure=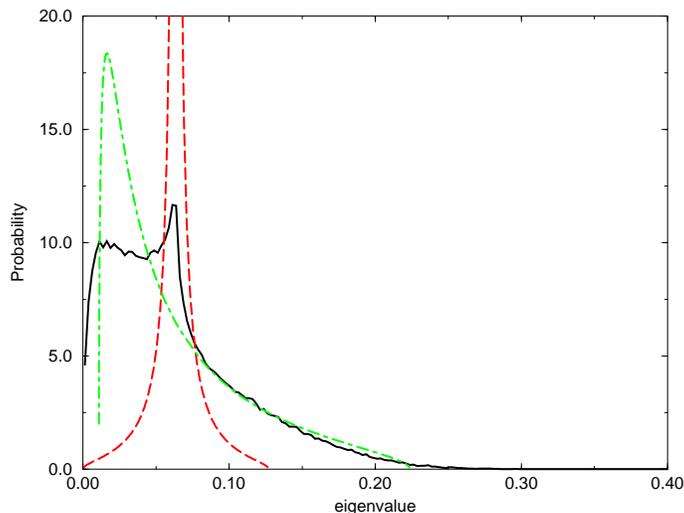,width=8 cm,angle=-90}
}}
\caption{ Density of eigenvalues of a Euclidean Random Matrix in three dimensions,
density $\rho=0.1$.
The function $f$ is
$f(x)=(2\pi)^{-3/2}\exp(-x^2/2)$, and the matrix is defined from
eq.(\ref{pb}) with $u=0$. The full line is the result of a numerical simulation
with $N=800$ points, averaged over $200$ samples. The dashed (red) line
is the result from the low density expansion. The dash-dotted (green) line is the result 
from the Gaussian variational approximation (RPA) to the field theory.}
\label{fig3}
\end{figure}

In fig.\ref{fig3}, we plot the obtained spectrum,
averaged over 200 realizations, for $N=800$ points at  density $\rho=0.1$
(We have checked that there is no substantial variation of the plot when going from $N=600$
to $N=800$). Also
shown are the low density approximation (\ref{lowrho}), and the result from
the variational approximation. We see from fig.(\ref{fig3}) that this
value of $\rho=0.1$ is more in the low density regime, and in particular there
exists a  peak around $\lambda=f(0)$ due to the isolated 
clusters containing small number of points.
The
variational approximation gives the main orders of magnitude
of the distribution, but it is not able to reproduce the details of the
spectrum, in particular the peak due to small clusters.
On the other hand the leading low density
approximation, which is not normalizable,  gives a poor approximation at 
this intermediate density.

\begin{figure}
\centerline{\hbox{
\epsfig{figure=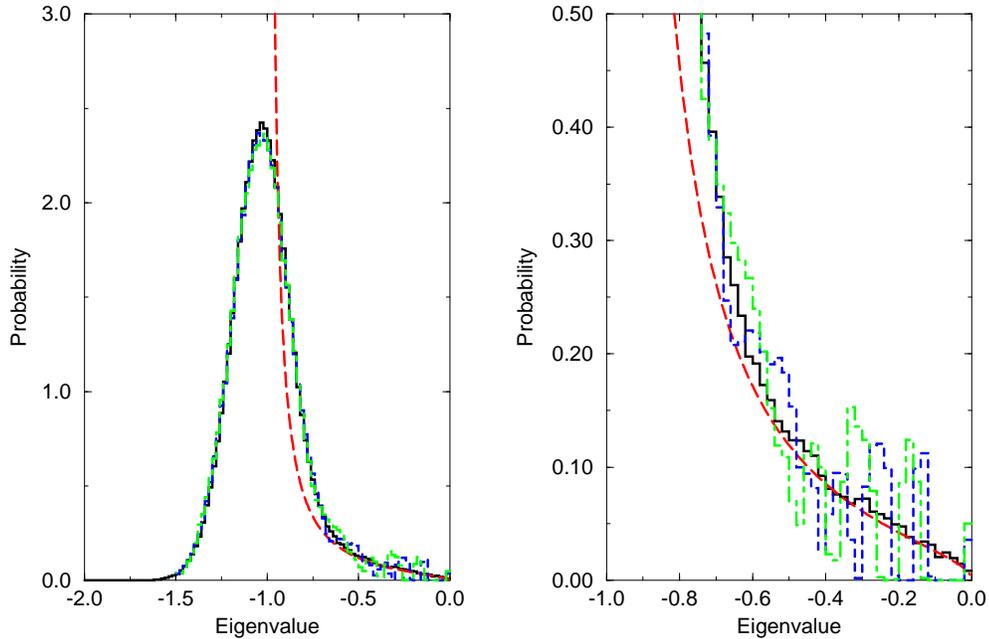,width=10 cm,angle=-90}
}}
\bigskip
\bigskip
\caption{ Density of eigenvalues of a Euclidean Random Matrix in three dimensions,
density $\rho=1$. The function $f$ is
$f(x)=(2\pi)^{-3/2}\exp(-x^2/2)$, and the matrix is defined from
eq.(\ref{pb}) with $u=1$. The right hand 
graph gives an enlargement of the spectrum at small eigenvalues.
The full line is the result of a numerical simulation
with quenched-twisted boundary conditions (see text), with
$N=500$ points, averaged over $500$ samples. 
The dashed (red) line is the result from the high density expansion.
The dash-dotted (green) line is the result of a simulation
with untwisted periodic boundary conditions, with $N=1000$
points averaged over $50$ samples. The dotted
 (blue) line is the result of a similar simulation, but with $N=1400$
 points.
}
\label{fig4}
\end{figure}

We now turn to the $u=1$ case.
We  have run some simulation in $d=3$ with the same Gaussian $f$ of width one,
at density $\rho=1$.
As seen in fig.\ref{fig4}, the direct simulations with periodic boundary conditions, 
and sizes $N=1000$
to $1400$, show roughly the same density of states, with a very broad peak
around  $\lambda=-1=-\rho \tf(0)$. However they disagree in the tail of the spectrum near
to the zero eigenvalue. In this region the finite size effects are more pronounced
because of the gap opening at momenta less than $2 \pi (N/\rho)^{-1/3}$, which is rather large
for the accessible values of $N$. In order to get a more precise result, less sensitive
to the finite sizes, in this region of the spectrum, we have run some simulations
which deal with an infinite set of points. Starting from a given sample 
of $N$ points in a box of side $L=(N/\rho)^{1/3}$, one imagines buiding
up an  infinite cubic lattice, for which the unit cell is the original box. 
Bloch's theorem tells that there exist $N$ bands, and the
 eigenstates of the infinite system are
the products of a periodic function of period $L$ (in each direction) 
times a plane wave $e^{i p x_j}$. For a given value of the momentum
$p\in [-\pi/L,\pi/L]^3$, the $N$ eigenvalues are those of the modified matrix:
\be
M^{(p)}_{ij}=f\((d_{ij}\))e^{i p.\phi_{ij}} -
u \delta_{ij} \sum_k f\((d_{ik}\)) \ .
\ee
Here $d_{ij}$ is the distance between $x_i$ and $x_j$ with
periodic boundary conditions,  defined as the minimum over all translation
vectors $t$ (translations of length multiple of $L$ in each direction) 
of $|x_i - x_j -t|$.
The phase $\phi_{ij}$ is a d-dimensional vector. Its component in direction $\mu$
is related to the optimum translation $t$ through:  $\phi_{ij}^\mu= t^\mu/L$.
Given $p$, the problem is thus mapped to finding the spectrum
of a $L^d \times L^d$ matrix $M^{(p)}$,
with randomly twisted boundary conditions. The resulting matrix 
is hermitian and can be diagonalized by standard library routines. 
We have averaged the spectra over many samples, choosing
for each sample one random Bloch momentum. This method \cite{twist} allows to 
access the small eigenvalue region of the spectra. As we see from fig.\ref{fig4},
this whole region is well approximated by the high density approximation 
(\ref{spec_highrho}), but this approximation fails to
reproduce the broad peak of the spectrum, as expected.

To summarize, we have introduced a family of random matrices with Euclidean correlations,
and developed for them a field theory representation, as well as some systematic
expansions giving some properties of the eigenvalue density. The specra can be rather 
well approximated in the simplest case $u=0$ using some variational
approximation to the field theory.  In the most interesting case $u=1$,
the situation is more complicated 
and we could get only the behaviour of the 
spectrum at small eigenvalues, using the high density expansion.
It would certainly be interesting to
generalize the variational approximation 
in order to treat also this $u=1$ case.

\end{document}